\documentclass[preprint,12pt]{elsarticle}

\usepackage{amssymb}
\journal{Physica C}

\begin{document}

\begin{frontmatter}



\title{Superconductivity of magnesium diboride}

\author{Sergey L. Bud'ko\corref{cor1}}
\ead{budko@ameslab.gov}

\author{Paul C. Canfield\corref{cor2}}
\address{Ames Laboratory US DOE and Department of Physics and Astronomy, Iowa State University, Ames, Iowa 50011, USA}

\cortext[cor1]{Corresponding author, phone: +1-515-294-3986; FAX: +1-515-294-0689}

\begin{abstract}
Over the past 14 years MgB$_2$ has gone from a startling discovery to a promising, applied superconductor. In this article we present a brief overview of the synthesis and the basic superconducting properties of this remarkable compound. In particular, the effects of pressure, substitutions and neutron irradiation on superconducting properties are discussed.

\end{abstract}


\end{frontmatter}


\section{Introduction}
\label{intro}

\begin{figure}
\begin{center}
\includegraphics[angle=0,width=120mm]{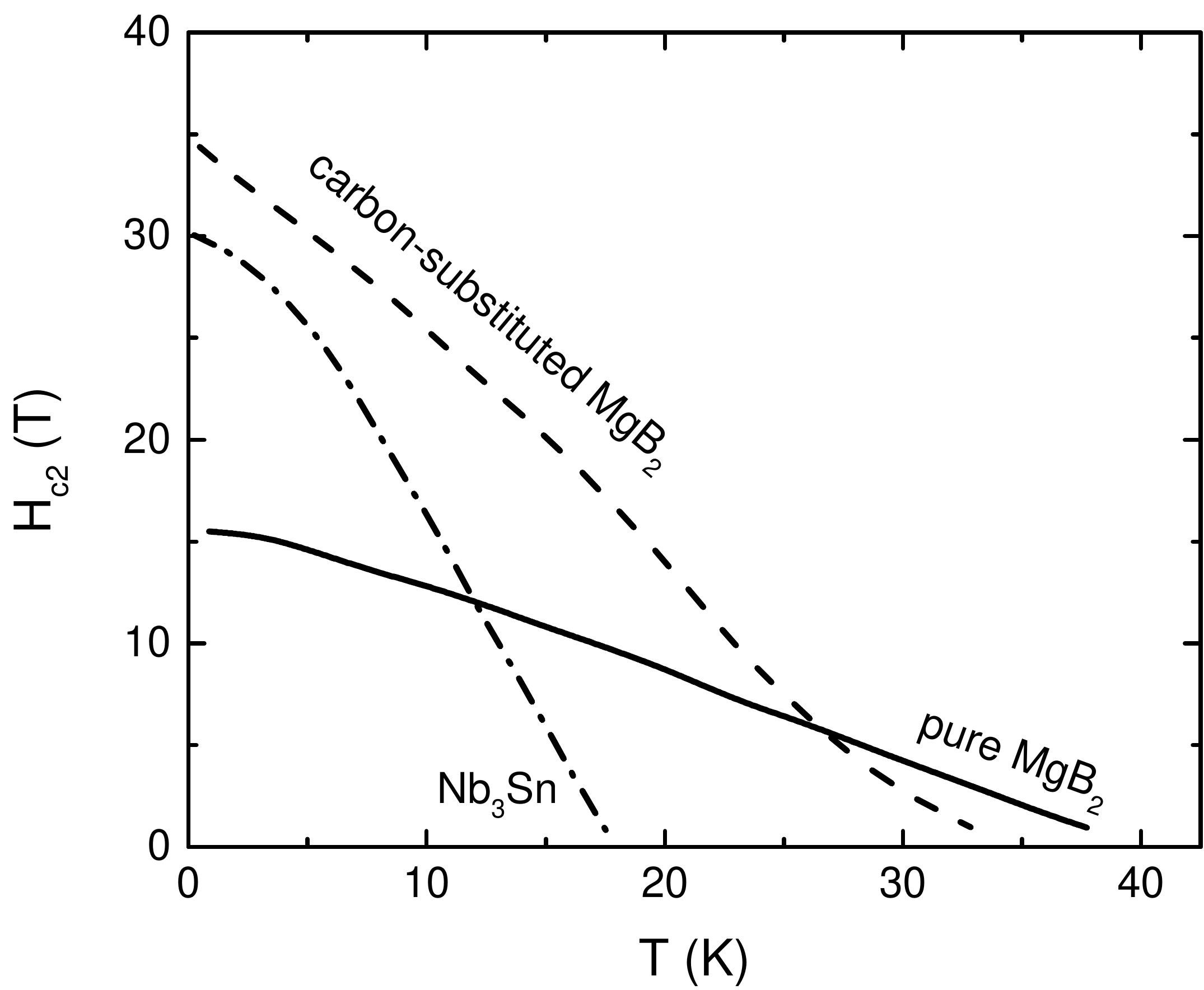}
\end{center}
\caption{Temperature-dependent upper critical field of pure and carbon-substituted ($\sim 5\%$ of C) MgB$_2$ in comparison with Nb$_3$Sn. (after Ref. \cite{can05a}.)}\label{Hc2com}
\end{figure}

On March 12, 2001 a large number of condensed matter physicists gathered for a technical session at the American Physical Society March Meeting in Seattle to share their results and thoughts on the newly discovered superconductivity in a simple binary compound, magnesium diboride (MgB$_2$). For somewhat nostalgic reasons (in memory of so-called "Woodstock of Physics" in 1987, where the discovery and first studies of superconducting cuprates were discussed) this session was quickly dubbed "Woodstock West". As in the case of superconducting cuprates, the discovery of superconductivity in MgB$_2$ \cite{nag01a} was serendipitous: the compound itself was known, \cite{jon54a} and even synthesized in single-crystalline form, \cite{fil67a} decades earlier. We can only conjecture that since this material appeared not to fulfill the preconcieved notion of what  high temperature superconductivity should look like, the proper experimental test for a possible superconducting ground state was not performed. The specific heat capacity of magnesium diboride was measured and tabulated from $\sim 18$ K to $\sim 305$ K, through the temperature that now we know to be its $T_c$, but no feature that could be associated with a transition was reported. \cite{swi57a} All in all, magnesium diboride was waiting almost 50 years for its moment, if not in "the sun", then at least in the cryostat. (As an aside, in a similar way, Ba- and Sr- doped La$_2$CuO$_4$ cuprates were synthesized and described, \cite{sha79a} but not cooled down below 78 K, almost a decade before the high temperature superconductivity in these materials was discovered. \cite{bed86a}). The discovery of superconductivity with $T_c \sim 40$ K in magnesium diboride caused initial excitement and hopes. Very quickly $H_{c2}(T)$ curves that exceeded those of Nb$_3$Sn were achieved  (Fig. \ref{Hc2com}) and a "MgB$_2$ zone of use" around 20 K was delineated. Indeed, in few years the physics of this material was considered understood and  applied research efforts took up the arduous task of furter improving critical current dencity, $J_c$, as well as wire development. 

In this brief review we summarize, mostly experimental, knowledge about superconductivity in magnesium diboride with more attention to the basic physical properties of bulk samples.   MgB$_2$ films will be mentioned only briefly, for details an interested reader can inquire several general and more specialized reviews.  \cite{buz01a,row02a,xix04a,nai04a,bri07a,xix09a} An interested reader will be able to find a number of in-depth discussions of specific topics related to superconductivity of magnesium diboride in two special issues of {\it Physica C} \cite{phc03a,phc07a} and in a number of reviews scattered in different sources, \cite{buz01a,iva03a,yan04a,eis07a,tom07a,xix08a,wil10a,mur11a} as well as more general, "popular science" reviews.\cite{can05a,can02a,can03a}

\section{Synthesis}
\label{synth}

MgB$_2$ forms in the hexagonal,  AlB$_2$ - type structure (space group: {\it P6/mmm}) with the lattice parameters $a = 3.0834$ \AA~and $c = 3.5213$ \AA. \cite{jon54a} In this structure the characteristic graphite-like two-dimensional layers formed by boron atoms sandwich the triangular Mg layers forming the structure similar to the intercalated graphite. Number of other metal - diborides are formed in the same structure. Details of chemical bonding in these materials are briefly surveyed by Spear \cite{spe76a}.

Bulk polycrystalline MgB$_2$ can readily synthesized by exposing solid boron to Mg vapor at elevated temperatures, \cite{bud01a} that is very close to the approach used in the initial publications on MgB$_2$. \cite{jon54a,swi57a} MgB$_2$ forms as a line compound \cite{mas90a} with possible Mg vacancies of, at most, less than one percent. \cite{hin02a} So to make stoichiometric MgB$_2$ one needs to mix magnesium and boron in 1 to 2 molar ratio and heat above the melting point of magnesium ($650^\circ$ C), although often higher temperatures, $\sim 950^\circ$ C are used. As MgB$_2$ is the most magnesium rich of all stable Mg - B binary phases, \cite{mas90a} to ensure complete reaction, it is possible to use excess of magnesium. This approach provides a form preserving method of turning boron objects (filaments, films, tapes) into MgB$_2$ objects with similar morphologies. \cite{can01a,shi01a,fin03a,fin03b} The properties of resultant MgB$_2$ (resistivity, $T_c$, upper critical field, etc.) can vary depending upon presence of impurities, in particular in boron. \cite{rib02a} 

Despite serious efforts of several groups there was no obvious success in growing MgB$_2$ single crystals of significant size (larger than few $\mu$m) at ambient pressure. The only viable technique for single crystal growth is  high pressure synthesis. Crystals with dimensions up to $1.5 \times 1 \times 0.1$  mm$^3$ and mass in excess of 200 $\mu$g can be grown under high pressure. The Mg - B phase diagram under pressure of 45 kbar \cite{tur03a} contains a eutectic at Mg : B $\approx 20 : 80$ with a temperature slightly above 550$^\circ$  C. This phase diagram presents the possibility of MgB$_2$ single crystal growth from solution at temperatures up to $\sim 2200^\circ$ C. Since in real experiments \cite{kar07a,lee07a} the synthesis is performed in a BN container which reacts with the melt at high temperatures (and pressures), the reaction in the ternary Mg - B - N   system should be considered. Synchrotron studies of the MgB$_2$ formation under pressure was reported in Ref. \cite{bar03a}, overviews of the MgB$_2$ single crystal growth under pressure were given in several publications. \cite{kar07a,lee07a} 

There were a number of attempts at partial replacement of Mg and B  in MgB$_2$ by different elements. To judge the veracity of such claims, a set of three criteria known as "Bob's rules" was suggested \cite{cav03a}: (1) second phases should not grow systematically in proportion with increasing dopant concentration; (2) a shift of the lattice parameters by more than three standard deviations in the series of doped samples should be seen; (3) properties of the superconductor should change on doping. The applicability of these criteria is not limited to substitution in MgB$_2$, but, in our opinion, should be considered in any study that involves chemical substitution. Of the attempts of substitution in MgB$_2$, only three are generally accepted as successful at this point: Mg(B$_{1-x}$C$_x$)$_2$, Mg$_{1-x}$Al$_x$B$_2$, and  Mg$_{1-x}$Mn$_x$B$_2$. \cite{kar07a,lee07a,cav03a,wil04a} These substitutions were achieved both in polycrystals and in single crystals and all cause a decrease in the superconducting transition temperature. It has to be mentioned that in the case of  Mg$_{1-x}$Al$_x$B$_2$ a two-phase region was reported for $0.1 < x < 0.25$. \cite{slu01a}

\section{Mechanism of superconductivity and electronic structure}
\label{mech}

\begin{figure}
\begin{center}
\includegraphics[angle=0,width=120mm]{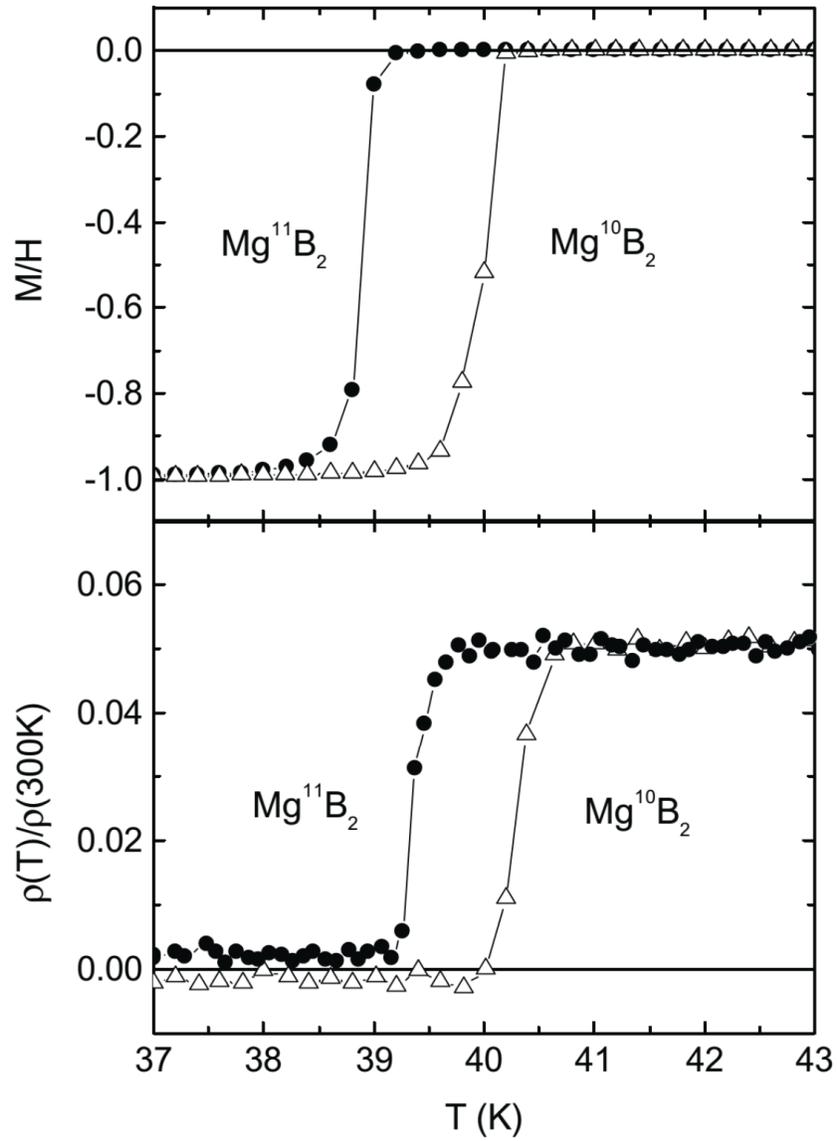}
\end{center}
\caption{Boron isotope effect as seen in magnetization (upper panel) and resistance (lower panel) (after Ref. \cite{bud01a}).}\label{iso}
\end{figure}

Experimentally, the mechanism of superconductivity in MgB$_2$ was addressed in one of the first publications: rather large difference in $T_c$ values, $\Delta T_c \approx 1$ K, was measured between Mg$^{10}$B$_2$ and  Mg$^{11}$B$_2$ (Fig. \ref{iso}). \cite{bud01a} The shift in $T_c$ associated with the Mg isotope effect is significantly smaller, $\sim 0.05 - 0.1$ K \cite{hin01a} The isotope effect coefficient, $\alpha$, is defined by $T_c \propto M^{-\alpha}$. The experimentally determined partial isotope effect coefficients for MgB$_2$ are $\alpha_B = 0.26 - 0.30$ and $\alpha_{Mg} = 0.02$ \cite{bud01a,hin01a}. Such a large isotope effect gives a strong indication of MgB$_2$ being a phonon mediated superconductor with boron phonon modes being dominant for superconductivity, that is there is a strong coupling between the conduction electrons and the optical $E_{2g}$ phonon, in which neighboring boron atoms move in opposite directions within the plane. \cite{hin03a} The total isotope effect in MgB$_2$ is still notably smaller than $\alpha = 1/2$ expected in the simple BCS case. This difference was addressed in several theoretical works \cite{yil01a,liu01a,cho02a,cho02b} and was suggested to be related to complex properties of MgB$_2$ including possible anharmonicity of the boron $E_{2g}$ phonon mode and multi-band superconductivity in this material. s-wave symmetry of the superconducting gap in MgB$_2$ has been inferred through measurements of $^{11}$B nuclear spin relaxation rate $1/T_1$ which was found to decrease exponentially in the superconducting state, revealing a small coherence peak just below $T_c$. \cite{kot01a}

\begin{figure}
\begin{center}
\includegraphics[angle=270,width=120mm]{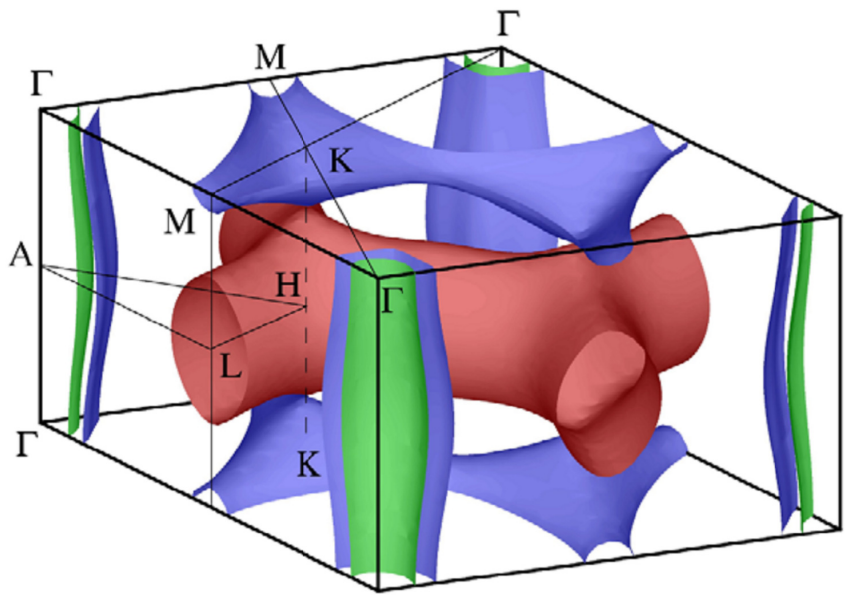}
\end{center}
\caption{Fermi surface of MgB$_2$ (from Ref. \cite{kor01a}).}\label{FS}
\end{figure}

Detailed band structure calculations for MgB$_2$ were reported in a number of publications that appeared soon after the discovery of superconductivity. \cite{anj01a,kor01a} $sp^2$ - hybridized boron atoms form $\sigma$ bonds with neighboring in-plane boron atoms. The boron $p_z$ orbitals form a 3-dimensional $\pi$ bonds by overlapping with both boron atoms within the plane and boron atoms in the adjacent planes. (The bonding and the terminology are very similar to what is found in benzene.) The in-plane $\sigma$ bonds give rise to quasi-2D cylindrical $\sigma$ bands, whereas $\pi$ bonds result in 3D planar honeycomb tubular networks (Fig. \ref{FS}). The extremal cross sections of the Fermi surface sheets experimentally observed in de Haas - van Alphen experiments \cite{car07a} are in a good agreement with the band structure calculations. The observed de Haas - van Alphen frequencies range between $\sim 500$ and $\sim 3000$ T, the effective masses are between $\sim 0.3 m_e$ and $\sim 1.2 m_e$. Comparison of the experiments and calculations point to the absence of any mass renormalization apart from that due to phonons. Additionally, the electron-phonon coupling in the $\sigma$ and $\pi$ bands appears to be very different, \cite{maz02a}  consistent with the basic view of two-band superconductivity in MgB$_2$ that will be briefly reviewed below.

\section{Two-gap superconductivity}
\label{two}

Despite being an s-wave, phonon mediated, BCS superconductor, MgB$_2$ became a playground for the study of several interesting aspects associated with complex superconductors. In this material the $E_{2g}$ phonon couples to both the $\sigma$ and $\pi$ bands resulting in two distinct superconducting gaps. Theoretically, two-band superconductivity was considered more than 50 years ago \cite{suh59a,mos59a}, experimentally, before MgB$_2$, only a few actual materials with possible two-band superconductivity were identified (see e.g. Ref. \cite{bin80a}), as a result the motivation for deeper studies of multi-gap superconductivity was limited. 

\begin{figure}
\begin{center}
\includegraphics[angle=270,width=120mm]{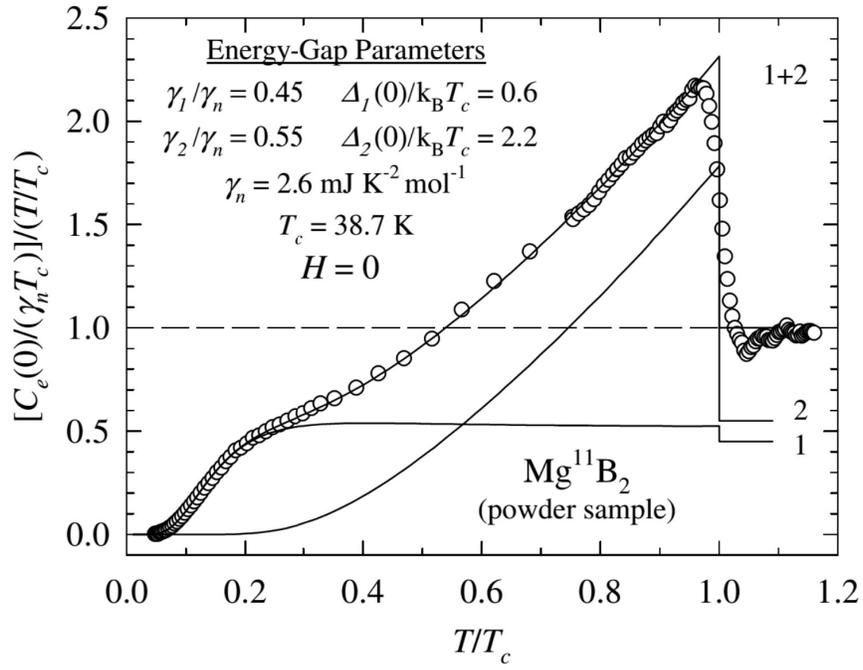}
\end{center}
\caption{Specific heat capacity of MgB$_2$ plotted as $[C_e(0)/\gamma_n T_c]/(T/T_c)$ vs. $T/T_c$. The solid curves are fits with a phenomenological two-gap model. Parameters obtained in the fit are listed in the plot (from Ref. \cite{fis03a}).}\label{Cp}
\end{figure}

Experimentally, the existence of two superconducting gaps  in MgB$_2$ was suggested in analysis of specific heat capacity data in superconducting state \cite{wan01a,bou01a}, but fitting the temperature dependent specific heat capacity data with a phenomenological two gap model \cite{bou01b} with meaningful parameters (Fig. \ref{Cp}), consistent with detailed band structure calculations, \cite{liu01a,cho02a,cho02b} was a significant advance towards proving two-gap superconductivity in MgB$_2$. \cite{fis03a} Additionally this analysis gave experimentalists a viable tool that since has been widely used in many novel superconductors. \cite{har14a} Theoretical discussions of the specific heat capacity and penetration depth in two-band superconductors can be found in Refs. \cite{mis05a,kog09a} and references therein.

Two-gap superconductivity was also inferred from measurements using Raman spectroscopy, \cite{che01a} photoemission, \cite{tsu01a} penetration depth, \cite{man02a} and tunneling. \cite{sza01a,giu01a}  High superconducting transition temperature, clean samples, well separated values of the superconducting gaps and almost equal partial Sommerfeld constants (similar relative weights) associated with $\sigma$ and $\pi$ bands made the experimental consequences of two - band superconductivity in MgB$_2$ relatively easy to observe.

\section{Basic physical properties and their anisotropies}
\label{basic}

\begin{figure}
\begin{center}
\includegraphics[angle=0,width=120mm]{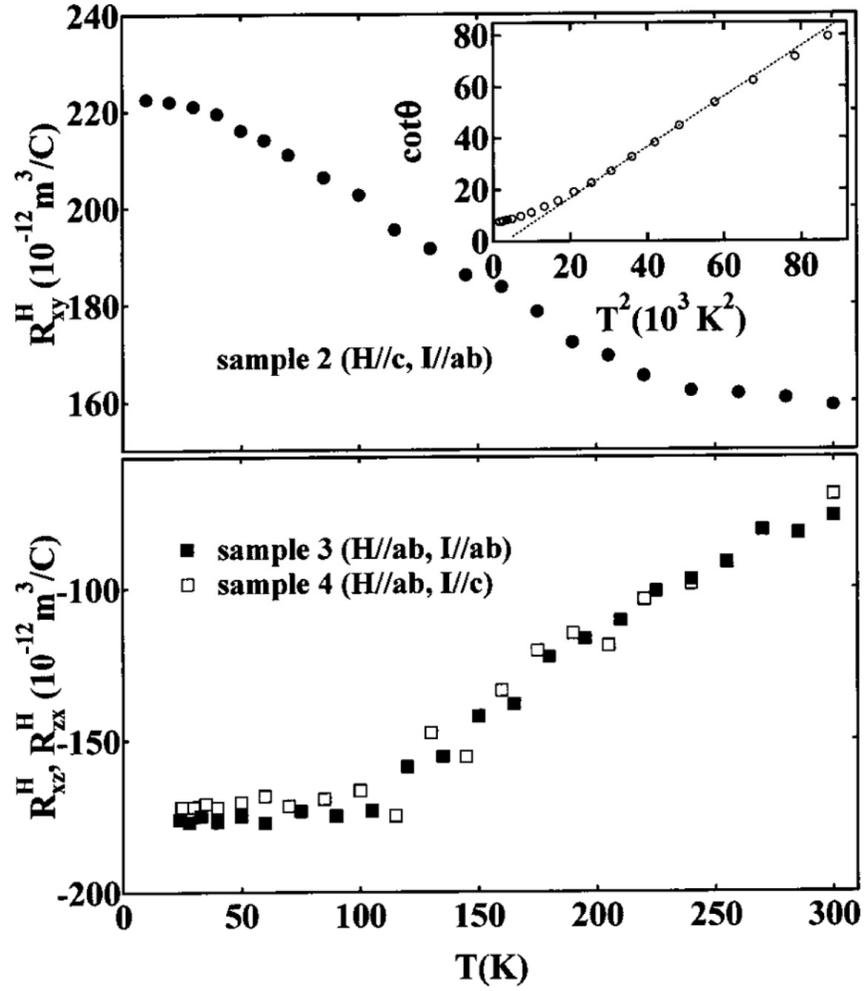}
\end{center}
\caption{Temperature dependent, normal state, in-plane and out-of-plane Hall constants in MgB$_2$ single crystals. Inset: temperature-dependent Hall angle, plotted as $cot~\theta_H$ vs. $T$ measured in 5 T applied field.  (from Ref. \cite{elt02a}).}\label{Hall}
\end{figure}

Basic physical properties as measured on high quality polycrystalline MgB$_2$ samples were reported almost immediately after discovery of its high $T_c$. From the specific heat capacity measurements \cite{bud01a} the Sommerfeld coefficient and the Debye temperature were estimated to be $\gamma = 3 \pm 1$ mJ/mol K$^2$ and $\Theta_D = 750 \pm 30$ K, respectively, with $\gamma$ value being relatively low, in the range common for normal metals, and $\Theta_D$ being rather high, exceeding by more than factor of two the values common for metals and intermetallics. It was easy to synthesize polycrystalline samples with residual resistivity ratio values, $RRR \equiv \rho(300$K$)/\rho(40$K$)$, of 20-25, \cite{can01a,fin01a} room temperature resistivity of $\approx~10~\mu \Omega$~cm and resistivity just above the superconducting transition, $\rho(40$K$) \approx~0.4~\mu \Omega$~cm. \cite{can01a} Anisotropic resistivity measurements on MgB$_2$ single crystals (hindered by the small size, in particular thickness, of the crystals) resulted in very similar temperature dependencies of $\rho_c$ and $\rho_{ab}$ (i.e. almost constant resistivity anisotropy, $\rho_c/\rho_{ab}$). In the initial study the room temperature values of resistivity were reported to be $\rho_{ab}(300$K$) \approx~6~\mu \Omega$~cm,   $\rho_c(300$K$) \approx~60 - 600~\mu \Omega$~cm,  that gives an estimate of $\rho_c/\rho_{ab} \sim 10 - 100$. \cite{mas02a} Later \cite{elt02a} slightly temperature dependent resistivity anisotropy $\rho_c/\rho_{ab} \sim 3.5 \pm 1$ was reported with the anisotropic residual resistivity values of 0.25 and 0.86 $\mu \Omega$~cm for $\rho_{ab}$ and $\rho_c$ respectively. The Hall constants were reported to be temperature dependent and of opposite signs for $H \| c$ and $H \| ab$ (Fig. \ref{Hall}), \cite{elt02a} indicating presence of both types of charge carriers and, thus, multi-band electronic structure of MgB$_2$. 

Initial thermodynamic and transport measurements on polycrystalline samples \cite{can01a,fin01a} allowed to roughly estimate (average) values of superconducting parameters of MgB$_2$: slope of the upper critical field and the thermodynamic critical field close to $T_c$,  $dH_c/dT \approx 0.44$ T/K
 and $dH_c/dT \approx 0.012$ T/K, respectively; the Ginzburg - Landau parameter, $\kappa = H_{c2}/(0.707 H_c) \approx 26$; the low temperature coherence length, $\xi_0 = \phi_0/(2 \pi H_{c2})^{1/2} \approx 5.2$ nm; the penetration depth, $\lambda = \xi_0/\kappa \approx 140$ nm. Using the estimate for residual resistivity of $\sim 0.4~\mu \Omega$ cm and calculated average Fermi velocity, $v_F \approx 4.8 \times 10^7$ cm/s \cite{kor01a} the electronic mean free path of $l \sim 60$ nm was inferred. This value is an order of magnitude higher than the superconducting coherence length, $l \gg \xi_0$, which meant that pure MgB$_2$ is deep in the clean limit of superconductivity.

\begin{figure}
\begin{center}
\includegraphics[angle=270,width=120mm]{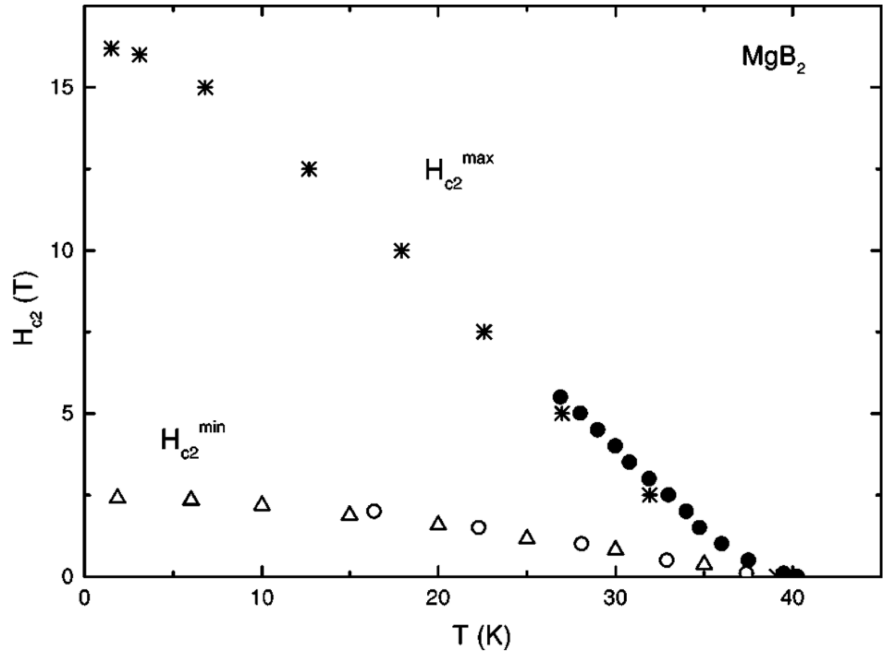}
\end{center}
\caption{Anisotropic upper critical field in MgB$_2$.  $ H_{c2}^{maxb}$ and $H_{c2}^{min}$ correspond to $ H_{c2}^{ab}$ and $H_{c2}^c$ respectively. (from Ref. \cite{bud02a}).}\label{Hc2}
\end{figure}

\begin{figure}
\begin{center}
\includegraphics[angle=270,width=120mm]{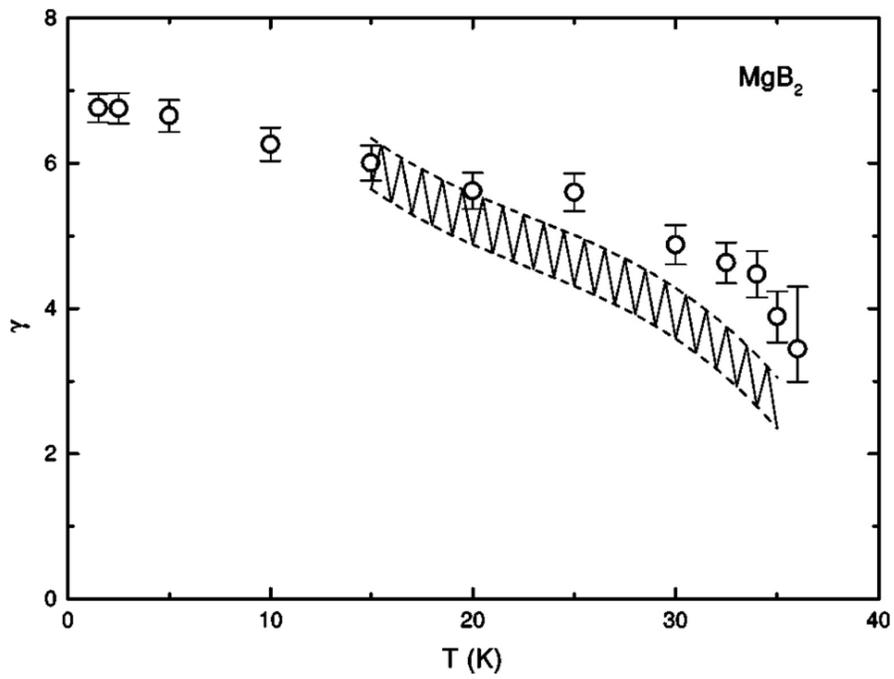}
\end{center}
\caption{Temperature dependent anisotropy of the upper critical field in MgB$_2$. Data obtained from torque measurements on single crystals \cite{ang02a} are shown as a hatched area between dashed lines for comparison.  (from Ref. \cite{bud02a}).}\label{ani}
\end{figure}

Resistivity measurements in high magnetic fields \cite{bud01b} confirmed that in the normal state MgB$_2$ behaves as good metal with high magnetoresistance that obeys the Kohler's rule and has a field dependence that can be described as $\Delta \rho(H)/\rho_0 \propto H^{\alpha}$ with $\alpha \sim 1.5$ (for high quality polycrystalline samples). Anisotropic upper critical field for MgB$_2$ data were inferred from a combination of magnetoresistance (that yields the maximum upper critical field) and magnetization measurements on clean polycrystalline samples, using the method developed in Ref. \cite{bud01c} (Fig. \ref{Hc2}). At low temperatures, the anisotropy ratio of $H_{c2}$ was determined to be $\gamma_H = H_{c2}^{ab}/H_{c2}^c \approx 6-7$. This value decreases down to $\gamma_H \approx 3-4$ close to $T_c$ (Fig. \ref{ani}). \cite{bud02a,ang02a}

\begin{figure}
\begin{center}
\includegraphics[angle=270,width=120mm]{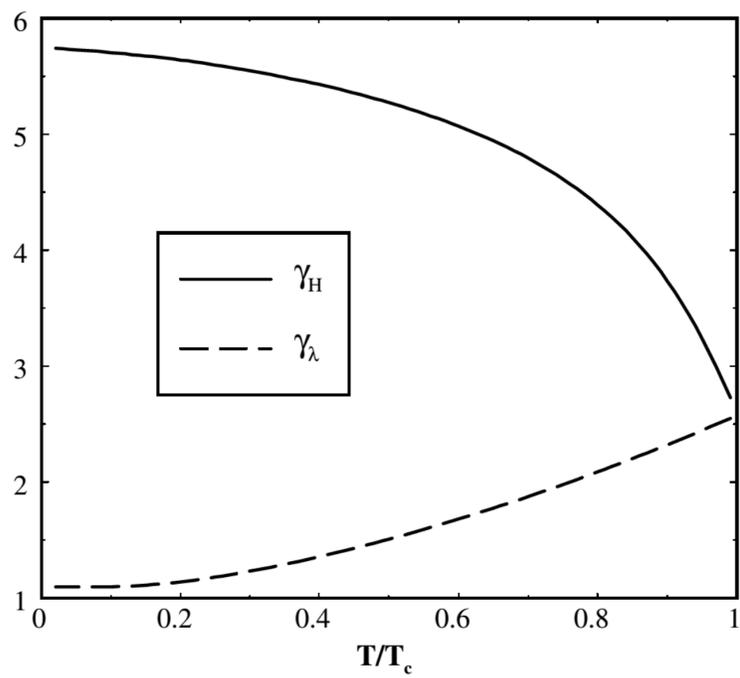}
\end{center}
\caption{Anisotropy of upper critical field (solid line) and London penetration depth (dashed line) as a function of reduced temperature $T/T_c$  (from Ref. \cite{kog03a}).}\label{the}
\end{figure}

Theoretical analysis of the superconducting anisotropy in MgB$_2$ within a weak coupling, two gaps, anisotropic s-wave model \cite{kog02a,kog02b,kog03a} revealed that the ratio of upper critical fields, $\gamma_H = H_{c2}^{ab}/H_{c2}^c$ is not equivalent to the ratio of the London penetration depths, $\gamma_{\lambda} = \lambda^c/\lambda^{ab}$, except at $T_c$. Using realistic parameters for MgB$_2$, it was shown that $\gamma_H$ increases with decreasing temperature, in agreement with experimental data, \cite{bud02a,ang02a} whereas $\gamma_{\lambda}$ decreases, to reach the value of $\approx 1.1$ at $T = 0$. (Fig. \ref{the}). Experimental studies of the flux line lattices  \cite{cub03a,cub03b}  and of anisotropy of the  upper  and lower critical fields, $H_{c2}$ \cite{bud02a,ang02a} and $H_{c1}$, \cite{lya04a} in MgB$_2$ are in agreement with the theoretical model. \cite{kog02a,kog02b,kog03a}  The concept of anisotropic gaps on anisotropic Fermi surfaces that yields two different anisotropy parameters, $\gamma_H$ and $\gamma_{\lambda}$, may lead to significant macroscopic consequences: \cite{kog03a}  the low temperature flux line lattice for $H \| ab$ is predicted to evolve from standard hexagonal in low fields to orthorhombic in increasing fields, with the disappearance of such behavior close to $T_c$. The difference between  $\gamma_H$ and $\gamma_{\lambda}$ should be taken into account e.g. when extracting anisotropy  data from torque measurements in tilted fields.  In general terms, beyond the particular example of MgB$_2$, it should be understood that a simple question "what is the anisotropy parameter of a superconductor?" may not have a unique answer. The proper question should specify the quantity of interest. Additionally, it should be mentioned, that all superconducting anisotropies might be suppressed by impurities. \cite{kog03a,kog04a}

\begin{figure}
\begin{center}
\includegraphics[angle=270,width=120mm]{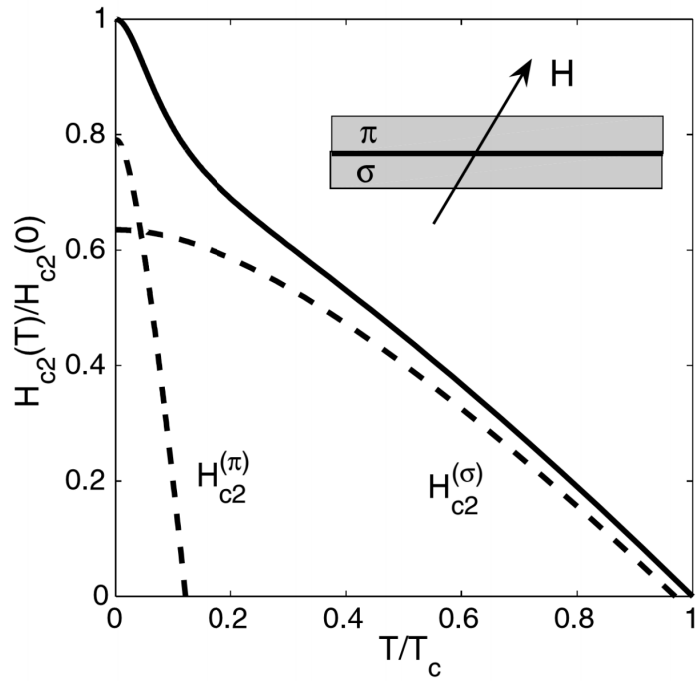}
\end{center}
\caption{Mechanism of upward curvature of $H_{c2}(T)$ illustrated by the bilayer toy model in the inset. Dashed curves: $H_{c2}(T)$ calculated for $\sigma$ and $\pi$ in the one gap dirty limit. Solid curve:  $H_{c2}(T)$ calculated  in the two gap dirty limit.  (from Ref. \cite{gur07a}).}\label{toy}
\end{figure}

Another important theoretical development followed the experimental observation of high $H_{c2}(T)$ curves with unusual curvature in dirty and carbon-doped MgB$_2$ thin films.\cite{gur04a} This behavior was analyzed within a model of two-gap superconductor in a dirty limit. \cite{gur03a,gur07a} Qualitatively the experimentally observed behavior can be understood in a simple bilayer model (Fig. \ref{toy}) where the physics of two band superconductivity is mapped onto a bilayer in which two films corresponding to $\sigma$ and $\pi$ bands are separated by a Josephson contact that models the inter-band scattering. \cite{gur07a} The global $H_{c2}(T)$ is then determined by the film (band) with the highest $H_{c2}$ even if $T_c^{(\sigma)}$ and $T_c^{(\pi)}$ are very different. This theoretical approach is reviewed in detail in Ref. \cite{gur07a} and has applications beyond superconductivity in MgB$_2$, possibly extending to iron based superconductors.

\section{Effects of pressure}
\label{pre}

The compressibility of MgB$_2$ is anisotropic, with $1/a_0 da/dP \approx - 1.9~10^{-3}$ GPa, $1/c_0 dc/dP \approx - 3.19~10^{-3}$ GPa and bulk modulus $B_0 \sim 150$ GPa. \cite{jor01a}, similar to $B_0 \sim 140$ GPa obtained in band structure calculations. \cite{loa01a}

$T_c$ of pure MgB$_2$ decreases under pressure with the initial pressure derivatives, $dT_c/dP$, ranging between $- 1.1$~K/GPa to $-1.6$~K/GPa. \cite{dee03a} It has been argued that the difference in the values of $dT_c/dP$ for nominally pure samples is caused by shear-stress effects in non-hydrostatic pressure media,rather than by the differences in the samples, with $dT_c/dP$ being close to  $- 1.1$~K/GPa in hydrostatic conditions. The pressure dependence of $T_c$ in MgB$_2$ was explained by strong pressure dependence of the phonon spectra. \cite{loa01a,med01a}

 Raman spectra, lattice parameters and $T_c$  of Mg$^{10}$B$_2$ and  Mg$^{11}$B$_2$ samples were measured under pressure up to $\sim 57$ GPa. \cite{gon01a,gon03a} The anharmonic character of the $E_{2g}$ Raman mode and anomalies in the pressure dependencies of the $E_{2g}$ Raman mode and $T_c$ at $\sim 15$ or $\sim 22$ GPA (boron isotope dependent) were interpreted as a result of a phonon - assisted Lifshitz electronic topological transition.

Comparative pressure studies were made on pure, Al - doped, carbon - doped and neutron damaged and annealed MgB$_2$ with the ambient pressure values of $T_c$ ranging from $\sim 8$ K to $\sim 39$ K. (Fig. \ref{pre}), \cite{bud05a} The observed difference in $dT_c/dP$ values for different samples with similar ambient pressure superconducting transition temperatures can be qualitatively understood by considering different levels of impurities / damage on boron layers (stronger effect in Mg(B$_{1-x}$C$_x$)$_2$, less in neutron damaged MgB$_2$ and virtually none for Mg$_{1-x}$Al$_x$B$_2$) and therefore on details of how the substitution or damage changes the phonon spectra.

Change of the upper critical field (for $H \| c$) under pressure up to 20.5 GPa for an MgB$_2$ single crystal was reported in Ref. \cite{sud04a}. The $T_c$ was reported to decrease by a factor of $\sim 2$ in this pressure range (consistent with other reports \cite{dee03a,gon03a}), whereas $H_{c2}^c(T \to 0)$ decreases by a factor of $\sim 4$  in the same range and the shape of the $H_{c2}(T)$ changes with increasing pressure. Based on these results it was argued \cite{sud04a} that in MgB$_2$ the electron-phonon coupling strength of the nearly two-dimensional $\sigma$ band, responsible for the high $T_c$, is more affected by pressure
than coupling in the $\pi$ band, and the hole doping of the $\sigma$ band decreases.

\section{Effects of chemical substitution}
\label{sub}

\begin{figure}
\begin{center}
\includegraphics[angle=270,width=120mm]{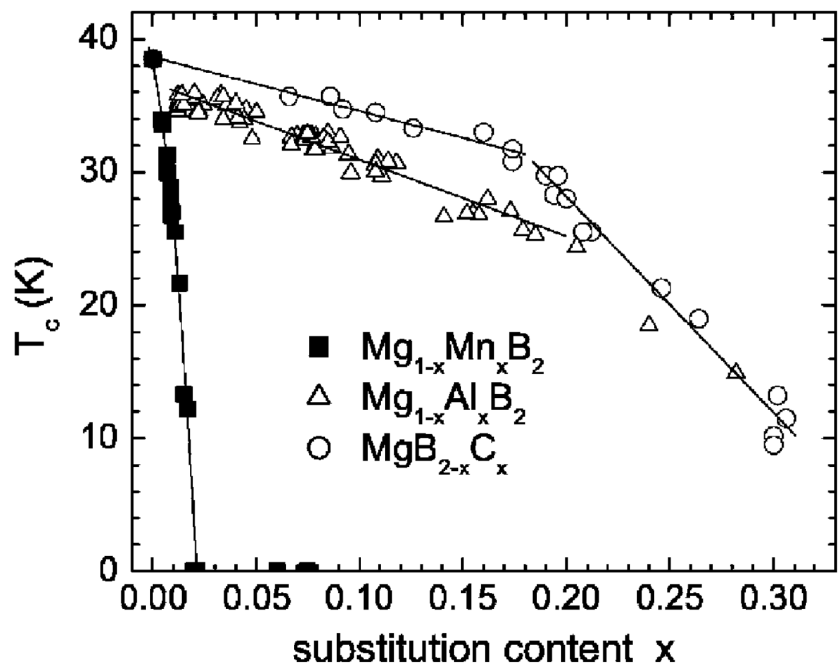}
\end{center}
\caption{$T_c$ as a function of substitution content, $x$, for  Mg$_{1-x}$Al$_x$B$_2$, Mg$_{1-x}$Mn$_x$B$_2$ and Mg(B$_{1-x}$C$_x$)$_2$ single crystals  (from Ref. \cite{rog06a}).}\label{Tc}
\end{figure}

Successful,  reliable and reproducible substitutions in MgB$_2$ are limited to three series: Mg$_{1-x}$Al$_x$B$_2$, Mg$_{1-x}$Mn$_x$B$_2$ and Mg(B$_{1-x}$C$_x$)$_2$ \cite{kar07a,lee07a,cav03a} (Fig. \ref{Tc}). In all three cases $T_c$ decreases with increase of doping, although the physics causing it is different.  In the case of Al substitution for Mg and C substitution for B the decrease of $T_c$ values can be understood  in terms of a band filling effect. Al and C are both electron dopants which reduce the number of holes at the top of the $\sigma$ bands together with a reduction of the electronic density of states. A simple scaling of the electron-phonon coupling constant  by the variation of the density of states as a function of electron doping was considered  sufficient to capture the experimentally observed behavior. \cite{kor05a} It should be hoted though that these two substitutions also introduce scattering centers which act in different ways (see discussion of $H_{c2}$ below).  

The rate of $T_c$ suppression in Mg$_{1-x}$Mn$_x$B$_2$ is significantly higher ($T_c$ is completely suppressed by 2\% on Mn doping). Experimental data suggest that  Mn ions are divalent and act as magnetic impurities, suppressing $T_c$ via a spin flip scattering mechanism. \cite{abr60a} $T_c(x)$ in the case of  Mg$_{1-x}$Mn$_x$B$_2$ follows the prediction of the Abrikosov - Gor'kov pair-breaking theory fairly well. \cite{rog06a} 

\begin{figure}
\begin{center}
\includegraphics[angle=270,width=120mm]{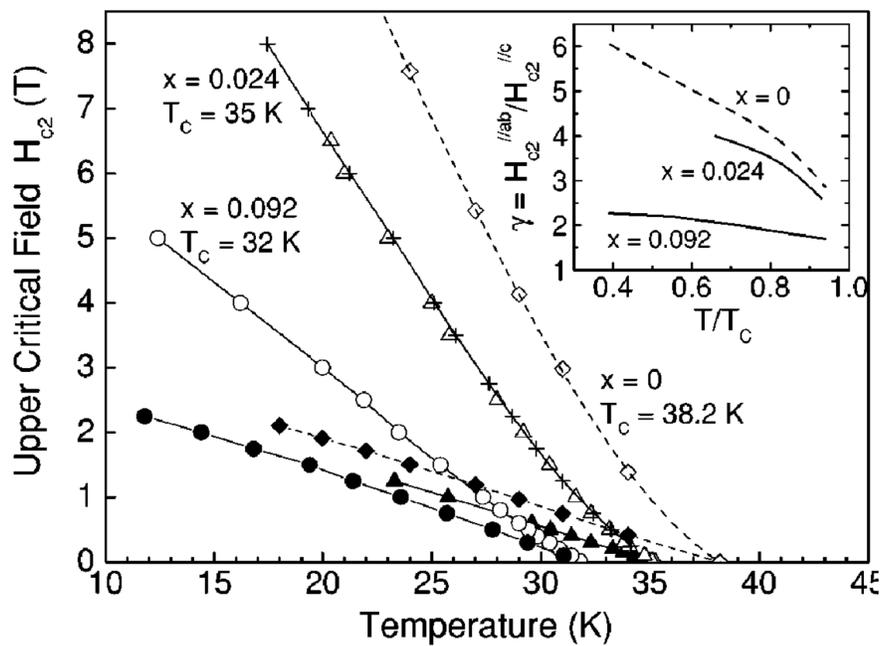}
\end{center}
\caption{Anisotropic $H_{c2}(T)$  for  Mg$_{1-x}$Al$_x$B$_2$   single crystals with different Al content. Inset: temperature - dependent $H_{c2}$ anisotropy for different $x$ - Al  (from Ref. \cite{kar05a}).}\label{Al}
\end{figure}

\begin{figure}
\begin{center}
\includegraphics[angle=270,width=120mm]{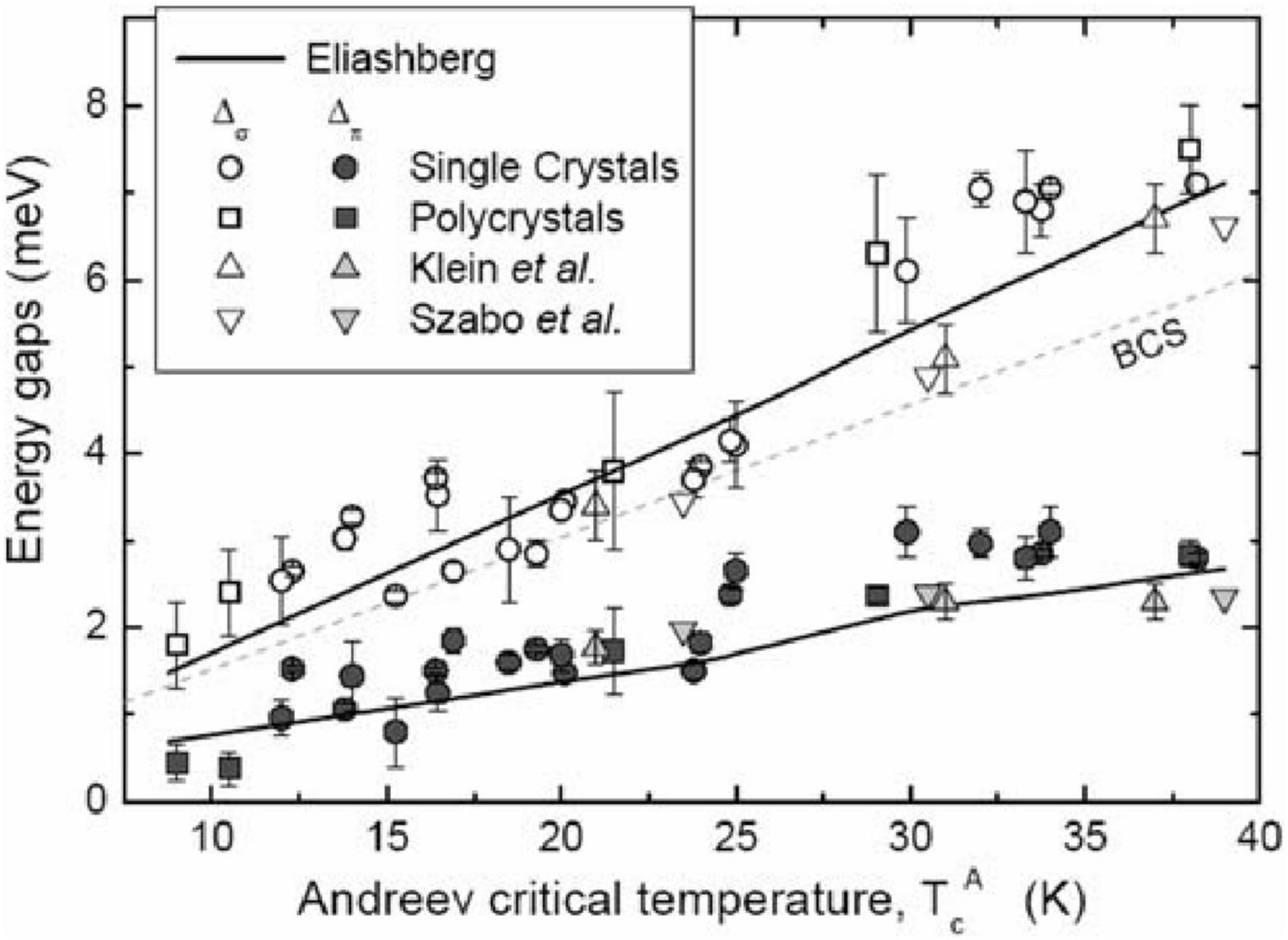}
\end{center}
\caption{Experimental values of superconducting energy gaps determined by point contact spectroscopy in  Mg$_{1-x}$Al$_x$B$_2$   plotted as a function of Andreev critical temperature. Solid lines - predictions of the two-band Eliashberg theory  (from Ref. \cite{gon07a}).}\label{AlG}
\end{figure}

In Mg$_{1-x}$Al$_x$B$_2$ the upper critical field, $H_{c2}(T)$, decreases with the decrease of $T_c$. A modest decrease in $H_{c2}^c$ is accompanied by a distinct decrease in  $H_{c2}^{ab}$, resulting in significant decrease of the $H_{c2}$ anisotropy (Fig. \ref{Al}). The superconducting gaps, as measured by point-contact spectroscopy, decrease with the increase of the level of Al - substitution (i.e. with the decrease of $T_c$), however no gap merging was observed up to Al concentration $\approx 0.32$) (i.e. down to $T_c \approx 9$~K) {Fig. \ref{AlG}}. This behavior of the superconducting gaps was analyzed within the two-gap, Eliashberg theory with the conclusion that band filling is the main effect of the Al substitution that is accompanied by some increase in inter-band scattering. \cite{gon07a}

\begin{figure}
\begin{center}
\includegraphics[angle=270,width=100mm]{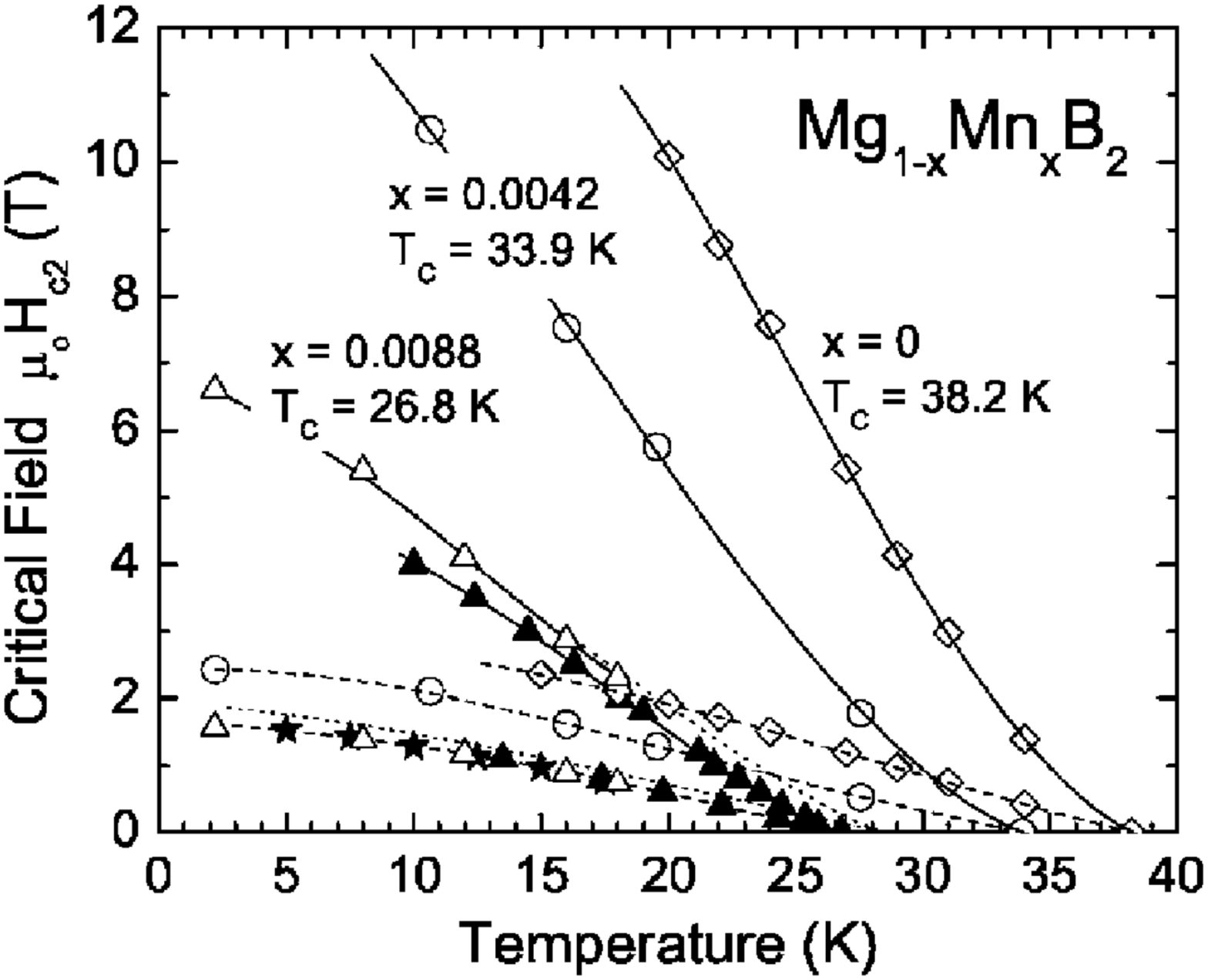}
\includegraphics[angle=270,width=100mm]{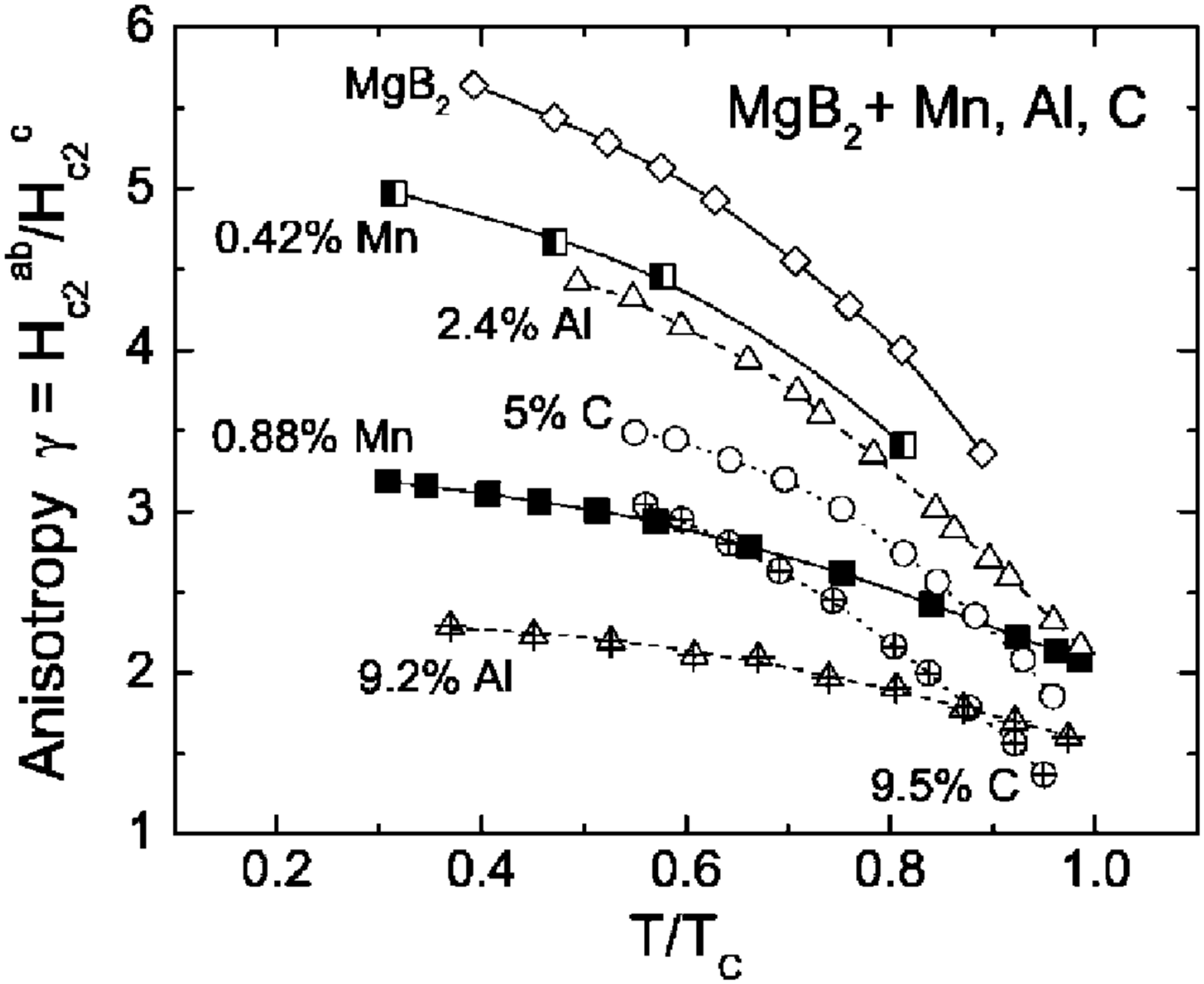}
\end{center}
\caption{Upper panel: anisotropic $H_{c2}(T)$ for  Mg$_{1-x}$Mn$_x$B$_2$ single crystals with different Mn content. Lower panel: $\gamma_H$ vs. reduced temperature for pure and substituted MgB$_2$ crystals  (from Ref. \cite{rog06a}).}\label{Mn}
\end{figure}

\begin{figure}
\begin{center}
\includegraphics[angle=270,width=120mm]{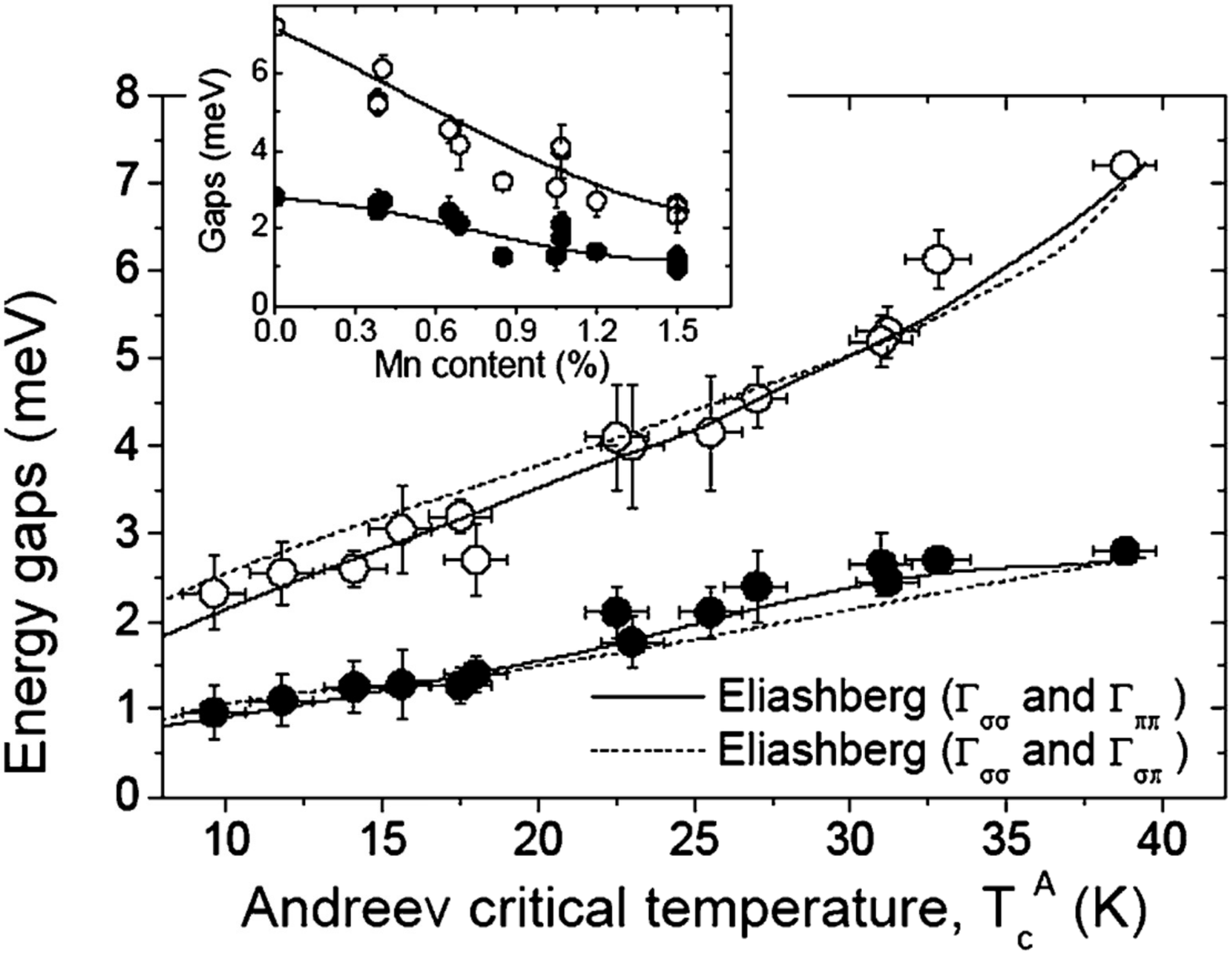}
\end{center}
\caption{Experimental values of superconducting energy gaps determined by point contact spectroscopy in  Mg$_{1-x}$Mn$_x$B$_2$   plotted as a function of Andreev critical temperature. Solid lines - predictions of the two-band Eliashberg theory. Inset: the same data plotted vs. Mn content  (from Ref. \cite{dag07a}).}\label{MnG}
\end{figure}

Despite the difference in the mechanisms of the $T_c$ suppression, there are similarities $H_{c2}$ and superconducting gaps evolution with substitution between Mg$_{1-x}$Al$_x$B$_2$ and Mg$_{1-x}$Mn$_x$B$_2$. In Mg$_{1-x}$Mn$_x$B$_2$ upper critical decreases with decrease of $T_c$ as well, with the changes in $H_{c2}^{ab}$ being more significant than the changes in the $H_{c2}^c$ (Fig. \ref{Mn}). As a consequence, the  $H_{c2}$ anisotropy decreases significantly with the Mn substitution. It was observed \cite{rog06a} that the temperature dependence of $H_{c2}$ and $\gamma_H$ for Mg$_{1-x}$Mn$_x$B$_2$ single crystals is similar to that for Mg$_{1-x}$Al$_x$B$_2$, provided the samples with similar $T_c$ were compared (Fig. \ref{Mn}), so  the value of $T_c$ controls both  $H_{c2}(T)$ and $\gamma_H(T)$ regardless of the mechanism responsible for the $T_c$ suppression. This observation probably suggests that both Al and Mn substitutions practically do not affect inter- and intra- band scattering (other than spin-flip scattering for Mn substitution) in any significant manner. The experimentally observed behavior of the superconducting gaps in Mg$_{1-x}$Mn$_x$B$_2$ single crystals as a function of the critical temperature (Fig. \ref{MnG} again is similar to the case of Mg$_{1-x}$Al$_x$B$_2$: the gaps decrease with increase of $x$ (i.e. decrease of $T_c$), but do not merge down to $T_c \approx 9$~K and can be analyzed within the two-gap Eliashberg theory. \cite{dag07a} \\

\begin{figure}
\begin{center}
\includegraphics[angle=270,width=70mm]{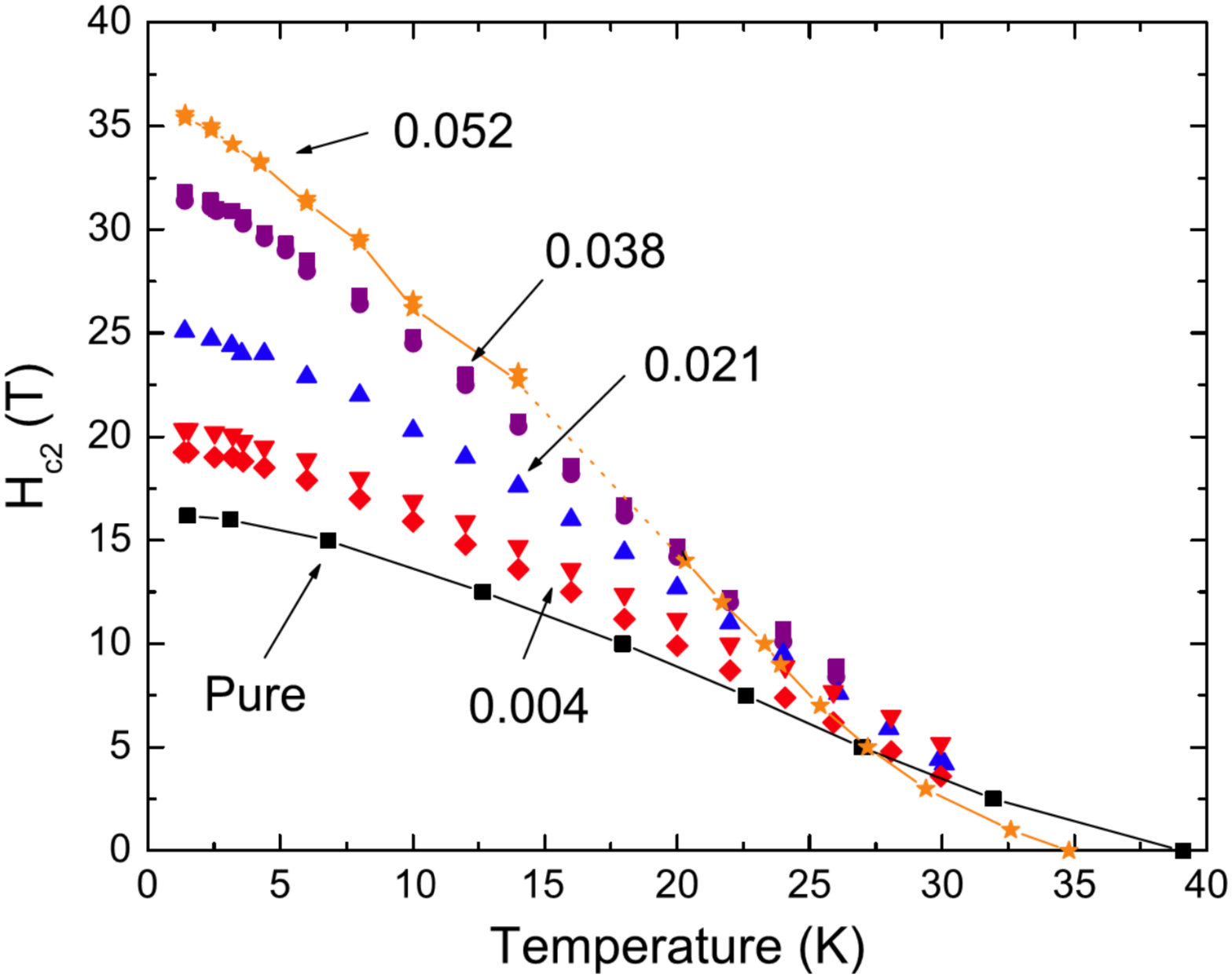}
\includegraphics[angle=270,width=70mm]{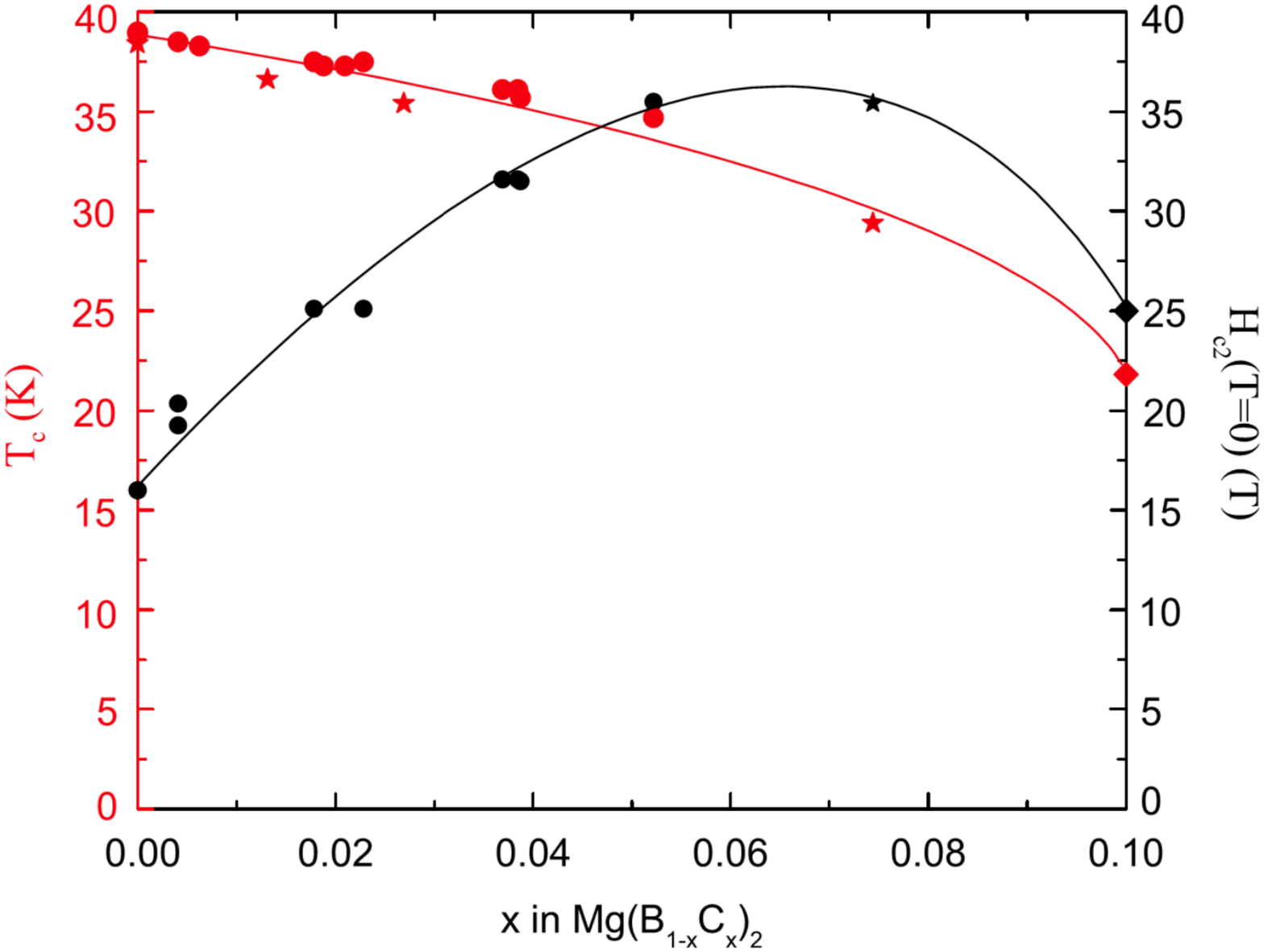}
\includegraphics[angle=270,width=70mm]{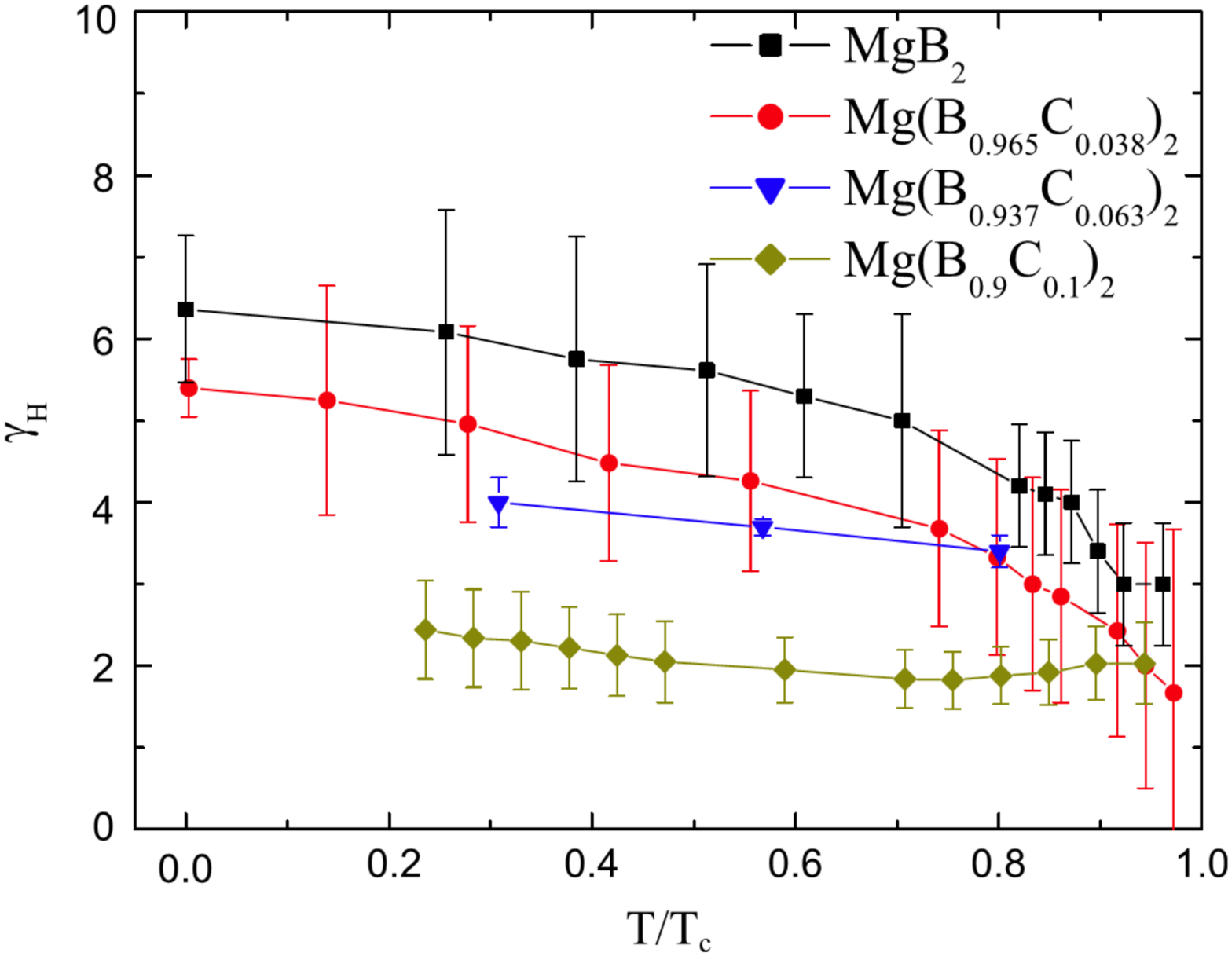}
\end{center}
\caption{Upper panel: $H_{c2}^{ab}(T)$ for Mg(B$_{1-x}$C$_x$)$_2$ with different C content. Middle panel: $T_c$ and $H_{c2}(T \to 0)$ as a function of C content in Mg(B$_{1-x}$C$_x$)$_2$.  (from Ref. \cite{wil04a,wil07a}) Lower panel: $\gamma_H$ vs. reduced temperature for pure and carbon substituted MgB$_2$ samples  (from Ref. \cite{ang05a}).}\label{C}
\end{figure}

\begin{figure}
\begin{center}
\includegraphics[angle=270,width=120mm]{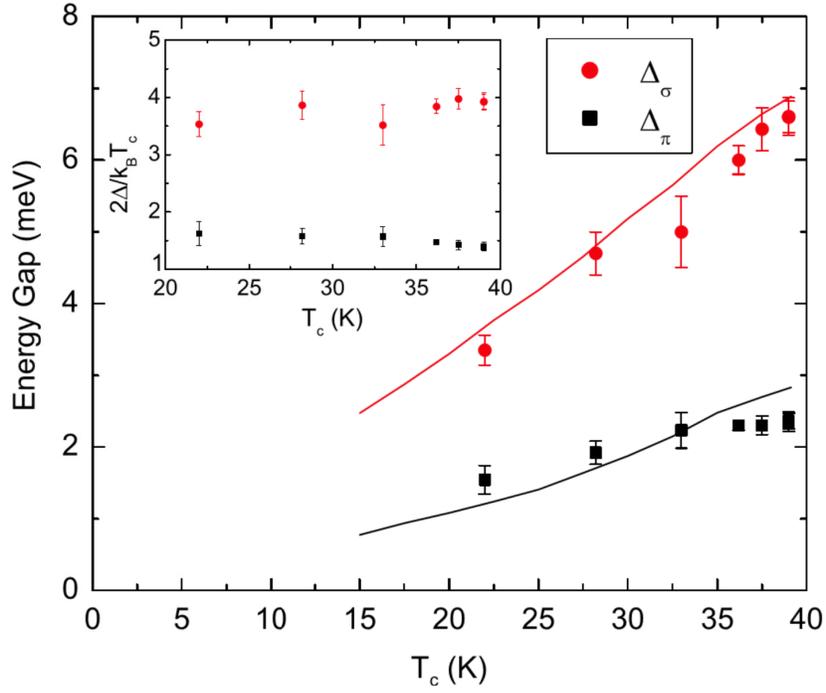}
\end{center}
\caption{Evolution of superconducting energy gaps determined by point contact spectroscopy in Mg(B$_{1-x}$C$_x$)$_2$   plotted as a function of superconducting critical temperature. Solid lines - theoretical predictions in the limit of no inter-band scattering. \cite{kor05b} Inset: ratio of the gap values to $T_c$  (from Ref. \cite{wil07a}).}\label{CG}
\end{figure}

In  Mg(B$_{1-x}$C$_x$)$_2$, the effects of substitution on the upper critical field were mostly studied on polycrystalline samples, resulting in determination of maximum upper critical field (e.g. $H_{c2}^{ab}$). In contrast to Al- and Mn- substitution, carbon substitution, while causing slight decrease in $T_c$, results in significant, more than two-fold, increase in $H_{c2}^{ab}$ (Fig. \ref{C}) \cite{wil04a,wil07a}, with $H_{c2}^{ab}(0)$ reaching more than 35 T for $x = 0.052$.   Using the approach for evaluation of $\gamma_H$ by measurements on polycrystalline samples \cite{bud01b} it was concluded the $H_{c2}$ anisotropy decreases significantly with C-substitution, down to temperature-independent $\gamma_H  \sim 2$ for Mg(B$_{0.9}$C$_{0.1}$)$_2$ (Fig. \ref{C}). \cite{ang05a}  As distinct from Al- and Mn- substitution, the evolution of $H_{c2}$ and $\gamma_H$ with C-substitution was suggested to result from increase of the intra-$\pi$-band scattering. \cite{sam06a} The superconducting gaps in  Mg(B$_{1-x}$C$_x$)$_2$ decrease as a function of the critical temperature. \cite{wil07a,sam05a,sam05b,gon07b,tsu05a} The measurements of superconducting gaps were performed for samples with superconducting critical temperature above 20 K, so the extrapolation of the observed behavior to lower $T_c$ samples is more ambiguous than in the case of Mg$_{1-x}$Al$_x$B$_2$ and Mg$_{1-x}$Mn$_x$B$_2$ and caused some discussion in the literature \cite{sam05a,gon07b}. The data in \cite{sam05a,sam05b,wil07a} (Fig. \ref{CG}) suggest that two gaps might merge somewhere in the range $0 < T_c < 10$~K, or will not merge at all. 

\section{Neutron-irradiated MgB$_2$ samples}
\label{neu}

\begin{figure}
\begin{center}
\includegraphics[angle=270,width=120mm]{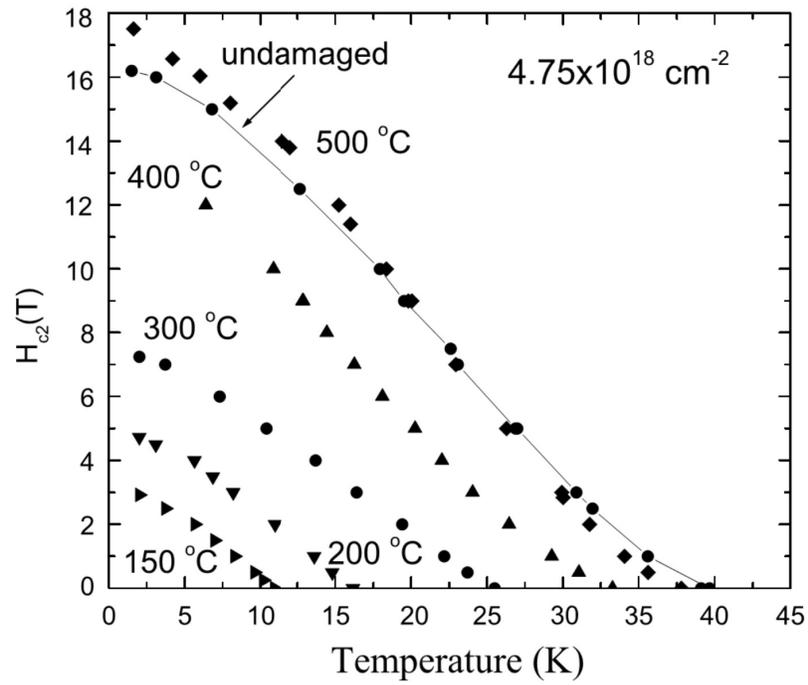}
\end{center}
\caption{Temperature dependent upper critical field for undamaged MgB$_2$ and the samples exposed to 4.75~10$^{18}$ cm$^{-2}$ fluence and annealed for 24 hours at different temperatures  (from Ref. \cite{wil06a}).}\label{Hc2N}
\end{figure}

Neutron irradiation (sometimes with subsequent annealing) was used to introduce defects in MgB$_2$ samples. \cite{wil07a,kar01a,eis05a,wil06a,dag06a,fer07a,put08a} It was shown that neutron irradiation suppresses $T_c$, increases resistivity and expands the unit cell.  By judicious choice of fluence, annealing temperature, and annealing time, the $T_c$ of damaged MgB$_2$ can be tuned to virtually any value between 0 and 39 K.  Although the samples and details of irradiation schedule differ in different publications, grossly speaking the results are consistent. The upper critical field decreases with increase of defects density (e.g. either larger fluence or lower post-irradiation annealing temperature) in a "Russian doll" pattern (Fig. \ref{Hc2N}) \cite{kar01a,wil06a}, except for rather small fluence or high post-irradiation annealing temperature, in which case although $T_c$ is 1 -2 K lower than in the undamaged sample, $H_{c2}(0)$ increases by several Tesla. \cite{wil07a,eis05a,wil06a}. This behavior possibly indicates that despite significant decrease of $T_c$ due to disorder caused by neutron damage, the samples can still be considered in a clean superconducting limit. \cite{kar01a} Although the data for $H_{c2}(T)$ anisotropy for neutron damaged samples are limited, \cite{put08a} it appears that the anisotropy decreases with increase of the number of defects. 

\begin{figure}
\begin{center}
\includegraphics[angle=270,width=120mm]{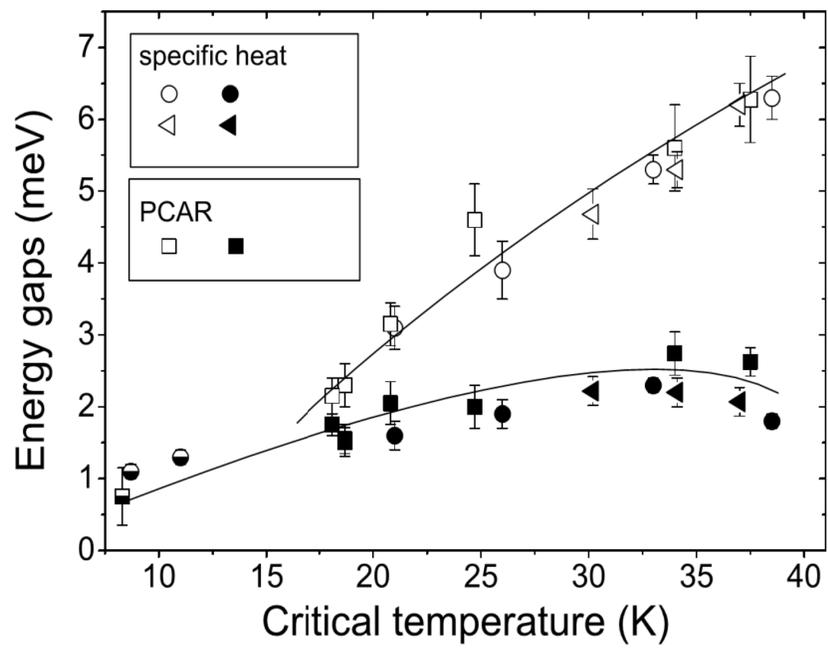}
\end{center}
\caption{Superconducting energy gaps in neutron-irradiated MgB$_2$ determined from specific heat capacity and  point contact spectroscopy measurements  (from Ref. \cite{fer07a}).}\label{NG}
\end{figure}

The effect of disorder caused by neutron irradiation on the superconducting gaps in MgB$_2$ has been studied using specific heat capacity and point contact spectroscopy \cite{fer07a}. The results suggested that the gap associated with the $\sigma$ band decreases with decrease of $T_c$ faster than the gap associated with the $\pi$ band, moreover, the two gaps coincide for the samples with $T_c < 12$ K (Fig. \ref{NG}). In the single gap region the gap value was reported to be lower than the BCS   value of $2 \Delta (0)/k_B T_c = 3.52$.

The modifications of the superconducting properties of MgB$_2$ by neutron damage were extensively compared with other perturbations used to modify Mg$_2$ properties. \cite{wil06a} Both pressure and neutron irradiation cause change of the lattice parameters, however the effects of these changes on $T_c$ are opposite. Using residual resistivity, $\rho(0)$ to parameterize $T_c$ gives similar trends for neutron damage and Al- doping, but not for C- doping, the values of $H_{c2}(T \to 0)$ as a function of $T_c$ are very different when neutron irradiation, Al- doping and C- doping are used. The changes in superconducting properties of neutron-irradiated samples were suggested to be primarily the result of an increase in the inter-band scattering and possible changes in the density of states at the Fermi level. \cite{wil06a} This comparison suggested that every perturbation used to modify superconducting properties of MgB$_2$ has its own, particular and complex, mixture of different mechanisms (inter-band scattering, intra-band scattering, band-specific change in density of states) that makes targeted tuning of the MgB$_2$ properties a difficult task.

\section{Summary and closing remarks}
\label{sum}

On one hand, the discovery of superconductivity in MgB$_2$ doubled the $T_c$ value for conventional phonon-mediated BCS superconductors and presented a clear example of i) two gap superconductivity and ii) low density of states, high temperature superconductivity. On the other hand, despite the initial hopes, discovery of high temperature superconductivity in MgB$_2$ was not followed by finding a family of related, interesting superconductors (although it did motivate broad searches for hints of superconductivity in wide classes of available materials \cite{fed14a}). Moreover, the allowed perturbations of MgB$_2$ are limited, with only C- doping in a  particular range of concentrations resulting in enhanced upper critical field.  This said, light, cheap and relatively isotropic MgB$_2$, modified to increase pinning and upper critical field, is considered as a viable material for MRI magnets and other applications. \cite{tom07a} On the basic research side, this material, overthrew the prejudices used in the search for new high temperature superconductors and brought along further and deeper experimental and theoretical understanding of two-band (broadly speaking, multi-band) superconductors, with some ideas and results being controversial until this time. \cite{mos09a,kog11a,dro14a}

\section*{Acknowledgments}
\label{ackno}

Work at the Ames Laboratory was supported by the US Department of Energy, Basic Energy Sciences, Division of Materials Sciences and Engineering under Contract No. DE-AC02-07CH11358.

\clearpage

\clearpage

\clearpage

\clearpage

\clearpage

\clearpage

\clearpage

\clearpage

\clearpage

\clearpage

\begin{figure}
\begin{center}
\includegraphics[angle=0,width=120mm]{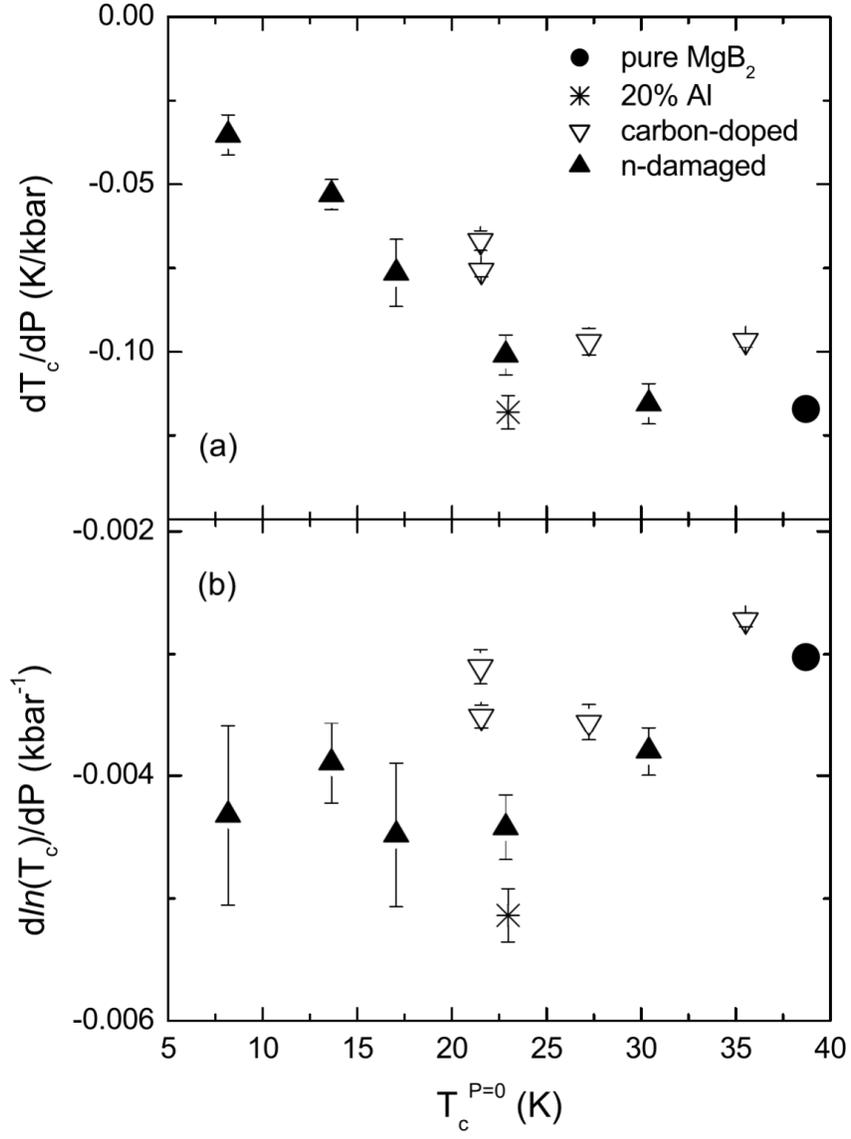}
\end{center}
\caption{Pressure derivatives, $dT_c/dP$ and $d~ln~T_c/dP$ as a function of the ambient pressure $T_c$ for pure, Al - doped, carbon - doped and neutron damaged and annealed MgB$_2$  (from Ref. \cite{bud05a}).}\label{pre}
\end{figure}

\clearpage

\clearpage

\clearpage

\clearpage

\clearpage

\clearpage

\clearpage

\clearpage

\clearpage

\end{document}